\documentstyle[bo99,epsfig]{article}
\def\etal{{\it et al.}}
\def\ref{\par\noindent\hangindent 15pt}

\title{Beat-Frequency Models of Kilohertz QPOs}
\author{M. Coleman Miller}
\affil{University of Maryland}

\begin{document}

\maketitle

\begin{abstract}
Kilohertz QPO sources are reasonably well-characterized
observationally, but many questions remain about the
theoretical framework for these sources and the
consequent implications of the observations for disk
physics, strong gravity, and dense matter. We
contrast the predictions and implications of the most
extensively studied class of kilohertz QPO models, the
beat-frequency models, with those of alternative classes
of models.  We also discuss the expected impact of
new observations of these sources with
satellites such as Chandra, XMM, and Constellation-X.
\end{abstract}

\section{Introduction}

Soon after the launch of the {\it Rossi} X-ray Timing Explorer
({\it RXTE}) in late 1995, observations with it of neutron star low-mass
X-ray binaries revealed kilohertz quasi-periodic brightness oscillations 
(QPOs) in the 
accretion-powered emission from many of these sources (see, e.g., 
van der Klis 2000 for a review).  These oscillations have high frequencies
(up to $\sim$1300~Hz [van Straaten et al. 2000], the highest-frequency 
astrophysical oscillations
ever observed), high amplitudes (up to 15\% rms in the 2-60~keV band of
the Proportional Counter Array on {\it RXTE}), and high coherences
(with quality factors $Q\equiv \nu/{\rm FWHM}>100$ in many cases), 
and often appear as two (but no more) high-frequency oscillations in
a single power density spectrum. 
Beat-frequency models (BFMs) were quickly proposed for this phenomenon 
(Strohmayer \etal\ 1996; Miller, Lamb, \& Psaltis 1998). In these models,
the higher-frequency of the two oscillations is attributed to the orbital
frequency of gas at some special radius near the star, and the lower-frequency
oscillation is a beat between this orbital frequency and the stellar spin
frequency.  These models are consistent with many of the trends evident
in the early data, including the approximate constancy of the frequency
difference between the two simultaneous
kilohertz oscillations and the close match in
four sources of this frequency difference with the stellar spin frequency
inferred from brightness oscillations during thermonuclear X-ray bursts.

It has, however, been established recently  that in several sources the
frequency difference is {\it not} constant, and indeed can vary by more
than 50~Hz.  Moreover, this variation is systematic: the higher the lower
peak frequency, the lower the frequency difference.  The explicability of
this behavior in the beat-frequency picture has direct bearing on  some
of the most important inferences drawn from the kilohertz QPOs.  For
example, only in BFMs is the leveling-off of the frequency of both
kilohertz QPOs possibly observed
from 4U~1820--30 (Zhang \etal\ 1998) a signature of the presence of the
innermost stable circular orbit (ISCO), a crucial prediction of
strong-gravity general relativity. Hence only in BFMs can one infer a
high gravitational mass $M>2.1M_\odot$ for this source, which constrains
strongly the equation of state of the high-density matter in
the core of neutron stars.

The changing separation frequency observed in several sources provided
part of the motivation for the development of other models of the
kilohertz QPOs, in particular the relativistic precession models (e.g.,
Stella \& Vietri 1998).
In these models the close match between the separation frequency and the
spin frequency inferred from burst brightness oscillations is a
coincidence, but they do predict the qualitative effect of a separation 
frequency that drops with increasing kilohertz QPO frequency.

Here we discuss the beat-frequency model in light of these new developments
and contrast it with alternate pictures.
In \S~2 we describe the observational trends that
motivated the development of beat-frequency models.  We then elaborate on these
models, in particular the sonic-point
beat-frequency model.  In \S~3 we discuss the evidence for a changing
difference frequency in the four sources Sco~X-1, 4U~1608--52, 4U~1728--34,
and 4U~1735--44.
We show that an aspect of the sonic-point beat-frequency model, included
in the dynamics but originally omitted from the frequency estimates,
naturally accommodates the changing difference frequency and can quantitatively
fit the data. 
Finally, in \S~4 we contrast some of the predictions of the beat-frequency
model with the predictions of the relativistic precession model, and
discuss analysis that might be done with current data to help discriminate
between the two interpretations.  We also explore the impact
of future observations, both with the upcoming generation of high
spectral resolution satellites (such as Chandra and XMM) and
with longer-term projects such as Constellation-X and a hypothetical 
high-area follow-on to {\it RXTE}.

\section{Motivation for Beat-Frequency Models}

As discussed in the introduction, soon after the discovery of kilohertz
QPOs it was established that these oscillations have (1)~high frequency,
(2)~high amplitude, and (3)~high coherence, and that there are always
two or fewer kilohertz QPOs in a given power density spectrum.  In
addition, the separation frequency appeared consistent with constant in
many sources and close to the spin frequency inferred from burst brightness
oscillations in the four sources where this could be tested.  It is now
known (see \S~3) that in several, and perhaps all, sources, the separation
frequency is {\it not} constant (although it is still close to the
inferred spin frequency), and in fact decreases systematically
with increasing lower peak frequency.  In \S~3 we discuss how this new
result may be interpreted within the beat-frequency model.

The high frequency indicates that the source of the brightness oscillations
is close to the neutron star.  A natural candidate for these oscillations
is the orbital frequency at some special radius.  Given that the burst
oscillation frequency is most convincingly interpreted as the stellar
spin frequency or its first overtone (see, e.g., Strohmayer \& Markwardt
1999), the close match of the separation
frequency with the inferred stellar spin frequency suggests a sideband
relation between the two simultaneous kilohertz QPOs.  The lack of a
third kilohertz QPO suggests a beat-frequency
relation, because most other mechanisms will produce both an
upper sideband and a lower sideband.  For example, amplitude modulation
of one frequency by another will produce two
sidebands of equal strength.
For all of these reasons, beat-frequency mechanisms are natural to propose
for the kilohertz QPO phenomenon.

The qualitative match of beat-frequency expectations with the data
is not enough to accept this model: it is also important to establish
a reason why a particular radius in the accretion disk would be
selected.  The high coherence of the oscillations requires, in this model,
that the radial range from which the oscillations are generated has a
fractional width less than $\sim 1/Q$, or less than $\sim$1\% in the
most stringent cases.  Moreover, the high amplitudes observed in some
sources indicates that the luminosity cannot be generated in just the
small range of radii where the frequency is determined.

For these reasons, the sonic-point beat-frequency model was proposed
(Miller, Lamb, \& Psaltis 1998).  In this model, the special radius
at which the frequencies are generated is where the inward radial
velocity of the gas in the disk increases rapidly with decreasing
radius.  As described in Miller, Lamb, \& Psaltis (1998), this rapid
increase in inward radial velocity is usually caused by radiation
drag, which removes angular momentum from the accreting gas and hence
causes it to spiral inwards.  If for some reason radiation drag is not
effective in a particular instance, as might happen if the accretion
rate is so high that the optical depth from the stellar surface is large,
then a similar rapid increase in velocity will nonetheless occur near
the ISCO, where the inspiral of gas will open
up simply because of gravitational effects.  The radial velocity typically
goes from subsonic to supersonic in this transition, and it is therefore
convenient to label it $r_{\rm sonic}$, the sonic point.  However, in
this model it is only the rapid change in velocity, and not the specific
fact that the gas crosses a sonic point, that is important.

In this model we assume that there are dense clumps
of gas orbiting at many radii near the star, and as we show in 
Miller, Lamb, \& Psaltis (1998) only the clumps near $r_{\rm sonic}$
will produce sharp brightness oscillations.  As gas streams from the
clump onto the star, it produces a bright pattern of impact that rotates
with the clump (see Miller, Lamb, \& Psaltis 1998), and hence a distant
observer sees a modulation of the flux from the system as the bright
impact spot passes into and out of view.  This occurs at approximately
the orbital frequency of the clumps at $r_{\rm sonic}$.  If the intensity of 
radiation
from the stellar surface is also modulated at the stellar spin frequency,
as it will be if a weak stellar magnetic field funnels extra matter 
towards one or two magnetic poles, then this variation in intensity
modifies the mass accretion rate from the clumps, and hence changes
the total luminosity from the impact spots.  As shown
in Miller, Lamb, \& Psaltis (1998), this creates an observed luminosity
modulation from the system at approximately the beat frequency between the 
orbital frequency of the clumps and the stellar spin frequency.  In this
model the luminosity in the oscillations is generated at the stellar
surface, where most of the gravitational energy is released, and hence
the oscillations can have high amplitudes.

We now demonstrate that in this model the slow inward drift of the clumps, 
included in the
dynamical calculations but not the frequency estimates
of Miller, Lamb, \& Psaltis (1998), generally produces a separation
frequency between the pair of kilohertz QPO peaks that is less than the
stellar spin frequency.

\section{Changing Difference Frequency}

\subsection{Observational Evidence}

By late 1996, observations of Sco~X-1 by van der Klis and colleagues
(van der Klis \etal\ 1997; see also M\'endez \& van der Klis 2000)
demonstrated that the separation frequency between the
kilohertz QPO peaks in this source is definitely not constant, and
in fact that the separation clearly decreases with increasing lower
peak frequency.
This is contrary to the simplest expectations of the
beat frequency model, in which the clumps generating the frequency are
orbiting at a constant radius and hence the separation is extremely
close to the stellar spin frequency.  

Many explanations for this behavior were discussed in the community.
Most of them centered on the apparently unique nature of Sco~X-1 in
this respect and hinged on other of its properties such as its very
high (near-Eddington) luminosity at the point when the change in the
difference frequency was most pronounced.  These explanations included
jet models (van der Klis \etal\ 1997) and other unpublished ideas that
involved effects such as the finite thickness of the accretion disk.
These models all had different physical
bases and even different mathematical models for the changing difference
frequency, but were all able to fit the relatively smooth $\Delta\nu$
vs. $\nu_{\rm lower}$ curve for Sco~X-1.

In the last year or so, however, careful analyses by M\'endez and
colleagues have shown that Sco~X-1 is {\it not} unique in having
a changing difference frequency.  Specifically, the low-luminosity
sources 4U~1608--52 (M\'endez \etal\ 1998), 4U~1728--34 (M\'endez \&
van der Klis 1999), and 4U~1735--44 (Ford \etal\ 1998)
also have difference frequencies that drop with increasing lower peak
frequency, and the separation frequency in 4U~1636--536 is slightly
but significantly less than the spin frequency inferred from burst
brightness oscillations (M\'endez, van der Klis, \& van Paradijs 1998).  
More generally, Psaltis \etal\  (1998)
showed that, within the observational uncertainties, all of the kilohertz
QPO sources have separation frequencies that are consistent with this
trend.  This is therefore clearly not an effect that requires a high
luminosity, and more general explanations must be sought.

\subsection{Explanation Within the Beat-Frequency Model}

As described in detail in Lamb \& Miller (2000), and as we now discuss,
in the sonic-point model 
the inward drift of the clumps tends to increase the frequency of the
lower kilohertz peak and decrease the frequency of the upper kilohertz
peak.  This produces a difference frequency that is generally less than the
stellar spin frequency, in agreement with observations.

If a clump were to orbit in
a perfect circle, then the inspiral time and phase traversed during inspiral
of gas from the clump to the stellar surface would be the same regardless
of when the gas separated from the clump.  In this case, consider two
successive ``beats", i.e., two successive maxima of the mass flow rate
from the clump.  These occur a time $\Delta T=1/(\nu_{\rm orb}-
\nu_{\rm spin})$ apart.  Because the inspiral time to the surface is the
same for each of the two beats, the arrival time at the surface is separated
by exactly that same $\Delta T$, and the observed frequency of the beat
is therefore just $1/\Delta T=\nu_{\rm orb}-\nu_{\rm spin}$.

Now consider clumps that are drifting in slowly.  Then the inspiral time
decreases as time progresses.  If we again think of two successive beats,
suppose that the inspiral time for the second beat is a small time
$\delta t$ less than the inspiral time for the first beat.  Then the
interval between releases of gas from the clump in the first and second
beats is still $\Delta T=1/(\nu_{\rm orb}-\nu_{\rm spin})$, but the
arrival times at the surface are now separated by $\Delta T-\delta t$.
Therefore the observed frequency, which depends on surface arrival
times because the surface is where the luminosity is generated, is
$1/(\Delta T-\delta t)$, which is
higher than it would have been without the inward drift.  This is
somewhat analogous to Doppler blueshifting.  Further analysis and
comparison with numerical simulations shows (Lamb \& Miller 2000) that
the upper peak frequency is also affected.  In fact, the inward drift
decreases the upper peak frequency, typically by a fractional amount 
that is $\sim$50\% of the fractional 
amount by which the lower peak frequency is
increased compared to the circular orbit approximation.  Both the
increase in the lower peak frequency and the decrease in the upper peak
frequency act to decrease the frequency separation.

As described in Lamb \& Miller (2000), the magnitude of the change
in the frequencies depends on the details of the inspiral, and in 
particular on how the total phase and time of the inspiral depend on
the radial location of the clump.  However, much of the physics is
captured in a simple model in which the frequency change depends only
on the ratio of two radial velocities: the velocity $v_{\rm clump}$
of the clump near
the sonic point and a characteristic velocity $v_{\rm gas}$
of the gas when it leaves
the clump (see Lamb \& Miller 2000 for more details).  The observed
frequencies are then approximately $\nu_{\rm low}=\nu_{\rm beat}/
(1-v_{\rm clump}/v_{\rm gas})$ and $\nu_{\rm high}=\nu_{\rm orb}
(1-{1\over 2}v_{\rm clump}/v_{\rm gas})$, where $\nu_{\rm beat}\equiv
\nu_{\rm orb}-\nu_{\rm spin}$ and $\nu_{\rm orb}$ is the orbital 
frequency at the sonic point.  Numerical simulations show that the
characteristic velocity $v_{\rm gas}$ of the gas is fairly 
insensitive to radius, and hence for simplicity we assume that
$v_{\rm gas}$ is independent of radius.

What remains is to produce a physical model for $v_{\rm clump}$ as a
function of radius.  The radial velocity of a clump
with angular momentum $J$ under the influence of a torque
$N$ is $v_r=N/(\partial J/\partial r)$.
In a Schwarzschild spacetime, for clumps whose mass does not depend
on the distance from the star,
\begin{equation}
\partial J/\partial r\propto {1\over 2} {r-6M\over{
\left(r-3M\right)^{3/2}}}\; ,
\end{equation}
where we use geometrized units in which $G\equiv c=1$.
The singularity at $r=6M$ in the equation for the radial velocity
is only apparent.  In reality, the velocity deviates from this simple
formula near the ISCO.  Numerical 
simulations show (Lamb \& Miller 2000) that for plausible torques
the radial velocity of the clump can be small enough that even near
the ISCO the inspiral time of the clumps is large,
and hence the coherence of the brightness oscillations can be as high
as observed.  The simulations also show that for small enough torques
the separation frequency can remain almost constant even near the
ISCO.  Therefore, in this model the observations of 4U~1820--30, 
which indicate an approximately constant separation frequency even 
when the kilohertz QPO frequencies may be leveling off, are still
consistent with the asymptotic frequency being approximately the
orbital frequency at the ISCO (Lamb \& Miller 2000).

As a particular model of the torque, assume that the clumps interact
only weakly with each other and with any background gas
in the accretion disk, so that the primary torque
is exerted by the stellar magnetic field.  The general form
of the torque at radius $r$ is then $N\sim B^2 Ar$,
where $A$ is the cross-sectional area of the clump and $B$ is the
strength of the stellar magnetic field at the location of the clump.  Assume
also that the clump mass is approximately
independent of the distance from the star, and that the clumps are in pressure
equilibrium with the external magnetic field.
Suppose that the clumps are kept at the local Compton
temperature, which is roughly independent of the radius.
Then $nkT\sim B^2\sim r^{-6}$ for a dipolar field, meaning
that the diameter of the clumps scales as $r^2$ and the
cross-sectional area scales as $r^4$.  Then the torque
scales as $N\sim r^{-1}$.

   This model can then be fit to the data, using the relations
between radial velocity and upper and lower QPO frequency
given above.  For a source such as Sco~X-1 which
has an unknown spin frequency, there are three parameters: the
stellar gravitational mass $M$, the spin frequency $\nu_{\rm spin}$, 
and the coefficient of the torque.  For a source such as
4U~1728--34 with a known $\nu_{\rm spin}$=364~Hz, there are two
free parameters.  Figure~1 shows the results of the fitting.  
It is encouraging that one physical
model can fit both of these sources as well as it does, especially
given the high quality of the data for Sco~X-1 (this model also fits
4U~1608--52 and 4U~1735--44).

\begin{figure}[t]
\begin{minipage}[t]{8.5cm}
\mbox{}\\
\centerline{\psfig{file=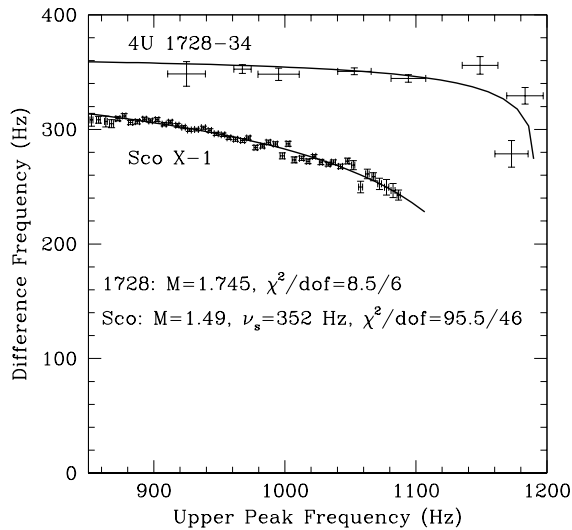, height=8cm, width=8cm}}
\end{minipage}
\begin{minipage}[t]{4cm}
\mbox{}\\
\caption[]{Difference frequency data and fit using
the drifting clump model described in this section.  The data for
4U~1728--34 and Sco~X-1 were kindly provided by Mariano M\'endez.
The spin frequency for 4U~1728--34 was fixed at 364~Hz to correspond
to the frequency of burst brightness oscillations in that source.}
\end{minipage}
\end{figure}

\section{Tests of Models With Current and Future Data}

With the immense archive of data accumulated with {\it RXTE},
there are a number of further analyses that could in principle
be performed to discriminate between models or test predictions
of models.  In this section we discuss such tests, particularly
as they might discriminate between beat-frequency models and
relativistic precession models, and also look to the future to
consider the qualitatively new types of observations that may
be available with the upcoming generation of X-ray satellites.

\subsection{Tests With Current Data}

An important difference between the predictions of the beat-frequency
model and the relativistic precession model concerns expectations for
the value of the stellar spin frequency.  In the beat-frequency
model it is expected that the stellar spin frequency is close to
(although slightly larger than) the separation frequency, and is
also very close to the burst oscillation frequency or half the burst
frequency, depending on the symmetry.  In contrast, as recently
emphasized by Psaltis \& Norman (2000), in the simplest version of 
the relativistic precession model the data on the lower frequency
horizontal branch type brightness oscillations combined with the
kilohertz oscillations requires that the stellar spin frequency
exceed $\sim$500~Hz in all sources, and be especially large (up to $\sim$
900~Hz) in the Z sources.  Simple estimates of corrections due to
fluid viscosity (Psaltis \& Norman 2000) suggest that these corrections
would increase further the required spin frequency.  The predictions
for $\nu_{\rm spin}$ therefore differ greatly in the two pictures, and
if a clear spin frequency is detected in the persistent, accretion-powered 
emission then this will have strong bearing on the modeling.  In
addition, if it happens that the spin frequency is not equal to the
burst oscillation frequency or half of it, this will overturn
the apparently compelling arguments that
burst brightness oscillations can only be 
explained via a rotational modulation mechanism.

The two models also differ in their predictions of the frequencies of
other, weak brightness oscillations.  In the relativistic precession
model, there is an upper sideband of the orbital frequency expected
(Psaltis \& Norman 2000).  The amplitude of this oscillation need not
be as great as the amplitude at the lower sideband; 
for example, if the underlying perturbations of the accretion disk cut 
off sharply at frequencies exceeding the observed orbital frequency then
the upper sideband could be weaker.
Nonetheless, the oscillation is expected to be there at some level.
In the beat frequency model there is an oscillation expected
at twice the observed lower peak frequency.  This is expected for 
several reasons.  For example, the waveform generated by the beat is
not expected to be perfectly sinusoidal, and hence overtones will be
generated.  In addition, if the radiation pattern rotating with the star,
which helps produce the beat, is itself not perfectly sinusoidal, then
it will generate weak overtones of the lower peak frequency.

The predicted frequencies discussed above depend only on the observed
frequencies.  If the observed frequencies are $\nu_{\rm low}$ and
$\nu_{\rm high}$, then the predicted frequency in the relativistic
precession model is $2\nu_{\rm high}-\nu_{\rm low}$ and the predicted
frequency in the beat frequency model is $2\nu_{\rm low}$.  The frequencies
will change with time for a given source, but these predictions are
ideally suited for testing by use of the shift-and-add method first
used by M\'endez \etal\ (1998) to discover a weak brightness oscillation
in 4U~1608--52.

\subsection{Tests With Future Data and Satellites}

The launch of a new generation of high spectral resolution satellites
such as Chandra and XMM opens up new ways to probe the strong
gravity and dense matter of neutron stars.  For example, suppose that
simultaneous observations of a kilohertz QPO source are performed with
{\it RXTE} and XMM.  If an Fe K$\alpha$ profile from the inner edge
of the nearly-circular flow is detected and characterized with XMM,
this profile gives the radius of that inner edge in units of the
gravitational mass of the neutron star.  If at the same time a pair of
kilohertz QPOs is detected with {\it RXTE}, then the frequency of the
upper peak, combined with the radius in units of the gravitational mass,
yields the mass of the star.  This is true independent of where the
inner edge is; that is, it need not be at the innermost stable circular
orbit.  Therefore, such simultaneous measurements could provide a clean way to
measure the gravitational masses of neutron stars in low-mass X-ray
binaries.

\begin{figure}[t]
\vskip-0.5truein
\begin{minipage}[t]{2.4truein}
\mbox{}\\
\psfig{file=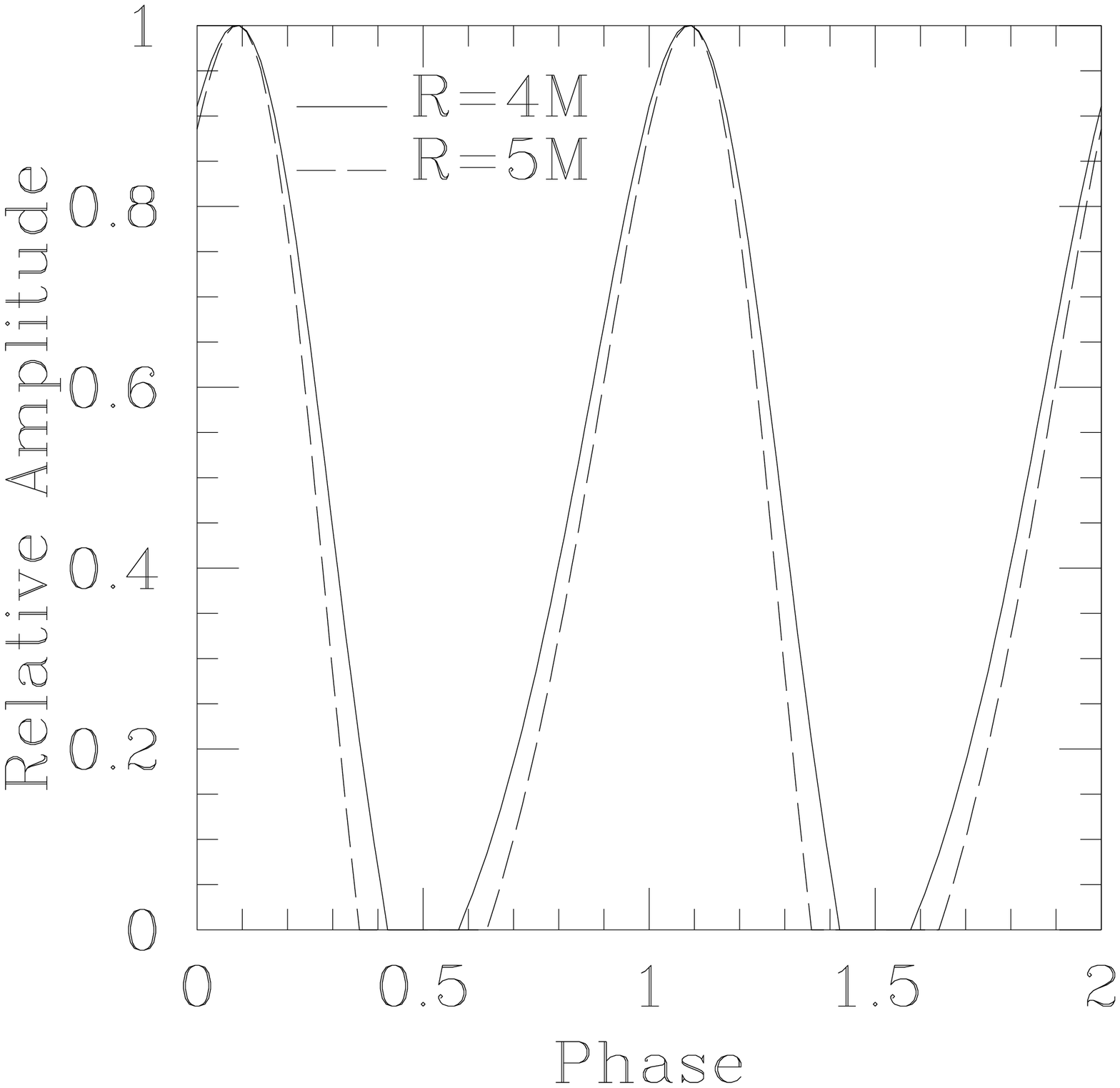, height=2.3truein,width=2.3truein}
\end{minipage}
\begin{minipage}[t]{2.4truein}
\mbox{}\\
\psfig{file=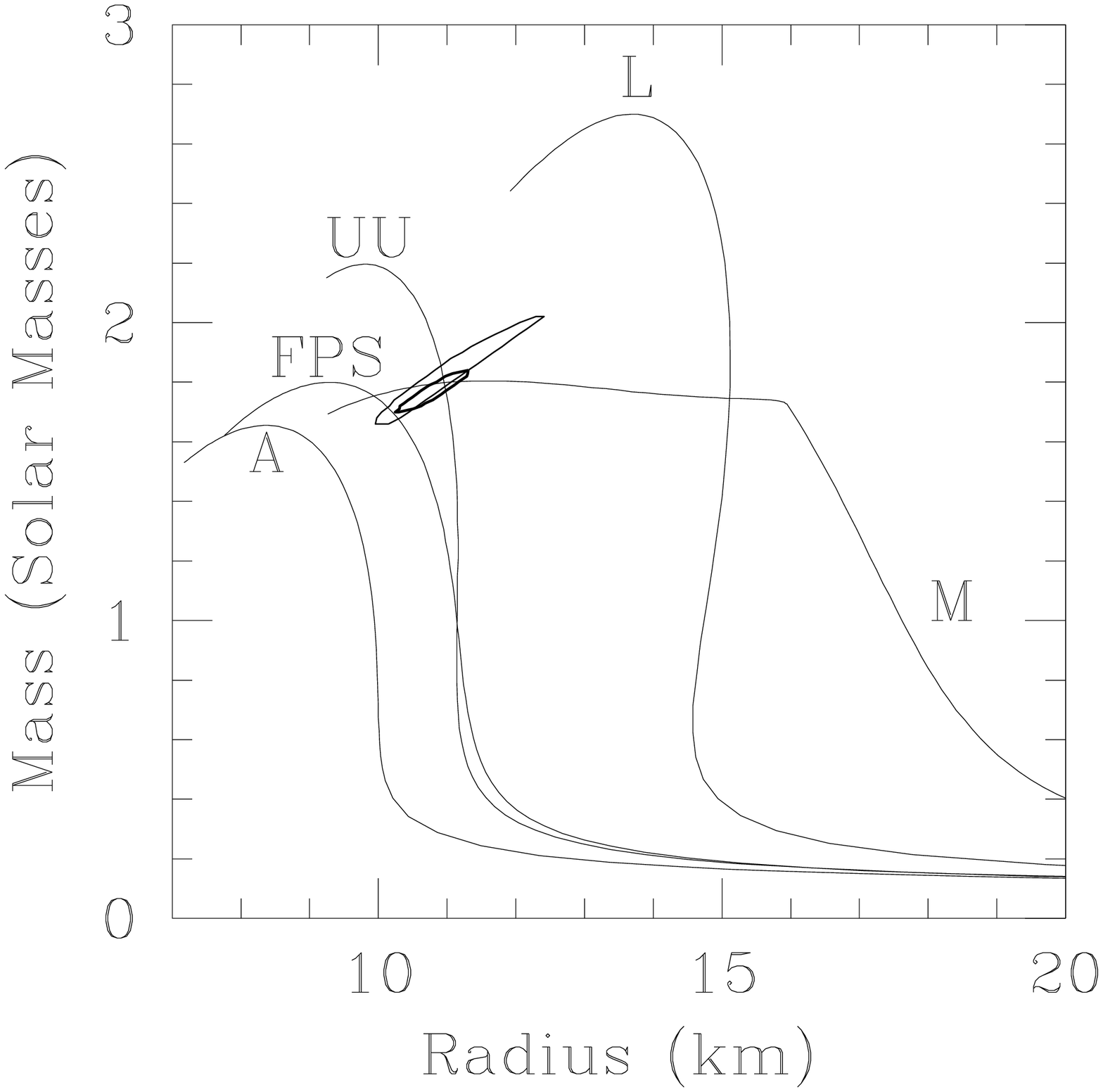, height=2.3truein,width=2.3truein}
\end{minipage}
\caption[]{\label{figlightcurve} (left panel)
Theoretical waveforms for burst brightness oscillations. Here we
assume a small bright spot on the rotational equator, as seen by
a distant observer in the rotational plane.  We show
simulated light curves over two cycles for a star with a
gravitational mass $M=1.8M_\odot$ and stellar radii
$R=4GM/c^2$ (solid curve) and $R=5GM/c^2$ (dashed curve), 
with a stellar spin frequency as seen at infinity of
364~Hz (the spin frequency of 4U~1728--34, the source most
frequently observed with {\it RXTE} to have burst brightness
oscillations). (right panel) Constraints on mass and
radius possible with waveform fitting of burst brightness
oscillations. We simulated the bolometric waveform from 
a 1.8 $M_\odot$ star rotating at 364~Hz, with high-density
equation of state UU, a small emitting spot on the rotational
equator, and no scattering after emission from the surface.
We then performed a likelihood analysis using this simulated
waveform.  We combined the constraints from five bursts,
assuming 5 seconds each of 5\% rms amplitude, and a 
flux typical of the bursts observed with {\it RXTE}.
The light outer contour shows the $1\sigma$ confidence region
expected from observations with Constellation-X, and the dark
inner contour shows the $1\sigma$ confidence region expected for
a hypothetical future 10~m$^2$ timing satellite.  The light solid
curves show the mass-radius relations given by different high-density
equations of state,
labeled as in Miller, Lamb, \& Psaltis (1998).}
\end{figure}

Future, high-area missions such as Constellation-X or a hypothetical
high-area follow-up timing mission to {\it RXTE} could provide even
more information, by allowing characterization of the waveforms of
the brightness oscillations seen in accretion-powered and burst-powered
emission.  An example is shown in the left panel of Figure~2, 
which shows theoretical waveforms for burst brightness oscillations.  
The curve for the more compact star is
broader, as is expected due to the extra gravitational light deflection.
In addition, the curve for the larger (less compact) star is more asymmetric,
due to the greater Doppler shifts from the surface rotation velocity.
The waveform therefore encodes information about both the mass and radius
of the star, meaning that repeated observations of bursts from a single source
can constrain the mass and radius tightly, with consequent constraints
on the equation of state of the high-density matter in the core of neutron
stars.

The right panel of Figure~2 shows an example of the constraints possible 
with Constellation-X
(outer contour, at 1$\sigma$) and a hypothetical 10~m$^2$ future timing 
instrument (inner contour, at 1$\sigma$).  Clearly, waveform fitting can
in principle yield very precise information about the mass and radius of
individual neutron stars, and therefore about the equation of state of 
matter at supranuclear densities.

High-area timing missions also will potentially allow us to do qualitatively
new types of tests of the strong-gravity predictions of general relativity.
For example, we have just discussed two independent ways of estimating the
gravitational mass of a neutron star: by combining Fe K$\alpha$ profiles
with kilohertz QPOs, and by waveform fitting of burst brightness oscillations.
If either of these is successful for an individual source, then we can 
predict precisely the orbital frequency at the innermost stable circular
orbit.  With high-area timing instruments we expect to see more cases like
4U~1820--30, in which there is leveling-off of the QPO frequency and thus
evidence for the ISCO.  The match of this asymptotic frequency with the
frequency predicted from the gravitational mass and general relativity
will provide us with unprecedented quantitative tests of general
relativity in strong gravity.  In conclusion, therefore, the continued
qualitative and quantitative agreement of the beat-frequency model with
observations of kilohertz QPOs has not only yielded important new
constraints on the equation of state of the dense matter in the core
of neutron stars and, possibly, the first direct evidence for unstable
orbits around neutron stars, a key prediction of general relativity.
It also indicates strongly that future observations of these sources,
especially with high-area timing missions, will allow us to continue to
make qualitative leaps in our observational understanding of strong
gravity and dense matter.

\begin{acknowledgements}
We thank Michiel van der Klis and Tod Strohmayer for discussions about
possible information from high-area timing missions.  This research was
supported in part by NASA ATP grant number NRA-98-03-ATP-028.
\end{acknowledgements}

\end{document}